\documentclass[aps,prc,twocolumn,superscriptaddress,showpacs]{revtex4}

\usepackage{graphicx}
\usepackage{dcolumn}
\usepackage{bm}
\usepackage{ulem}
\usepackage{color}

\begin{document}

\title{
  Direct study of the alpha-nucleus optical potential at astrophysical energies using the $^{64}$Zn(p,$\alpha$)$^{61}$Cu reaction
}

\author{Gy. Gy\"urky}
\email{gyurky@atomki.mta.hu}
\affiliation{Institute for Nuclear Research (Atomki), H-4001 Debrecen, Hungary}
\author{Zs. F\"ul\"op}
\affiliation{Institute for Nuclear Research (Atomki), H-4001 Debrecen, Hungary}

\author{Z. Hal\'asz}
\affiliation{Institute for Nuclear Research (Atomki), H-4001 Debrecen, Hungary}

\author{G.G. Kiss}
\affiliation{Institute for Nuclear Research (Atomki), H-4001 Debrecen, Hungary}

\author{T. Sz\"ucs}
\affiliation{Institute for Nuclear Research (Atomki), H-4001 Debrecen, Hungary}

\date{\today}

\begin{abstract}
In the model calculations of heavy element nucleosynthesis processes the nuclear reaction rates are taken from statistical model calculations which utilize various nuclear input parameters. It is found that in the case of reactions involving alpha particles the calculations bear a high uncertainty owing to the largely unknown low energy alpha-nucleus optical potential. Experiments are typically restricted to higher energies and therefore no direct astrophysical consequences can be drawn. In the present work a (p,$\alpha$) reaction is used for the first time to study the alpha-nucleus optical potential. The measured $^{64}$Zn(p,$\alpha$)$^{61}$Cu cross section is uniquely sensitive to the alpha-nucleus potential and the measurement covers the whole astrophysically relevant energy range. By the comparison to model calculations, direct evidence is provided for the incorrectness of global optical potentials used in astrophysical models.
\end{abstract}


\maketitle

Although chemical elements heavier than Iron represent only a tiny fraction of the matter of our world, the understanding of their stellar production mechanism remains a difficult problem of astrophysics. The bulk of the heavy elements is thought to be produced by neutron capture reactions in the s- and r-processes \cite{kap11,arn07}. While the s-process is relatively well known -- although some open problems still exist --, the r-process is still very poorly known regarding both the astrophysical site and the nuclear physics background. The synthesis of the so-called p-isotopes -- isotopes which are not produced by the s- and r-processes -- require further nucleosynthetic processes, like the $\gamma$-process \cite{rau13} or the rp-process \cite{sch98}. 

Common in the heavy element nucleosynthesis processes is that for their modeling huge reaction networks must be taken into account often including thousands of reactions. With the exception of the s-process these reactions mostly involve radioactive isotopes and therefore experimental information about these reactions is missing. Even at stable isotopes experimental data are very scarce owing to the tiny cross section at astrophysical energies. Consequently, reaction rates needed for the astrophysical network calculations are obtained from theoretical cross sections. In the relevant mass and energy range the dominant reaction mechanism is the compound nucleus formation and high level densities are encountered, the mostly used nuclear reaction theory is thus the Hauser-Feshbach statistical model. 

If the statistical model provides incorrect cross sections, then this may contribute to the failure of some astrophysical model calculations. This is found e.g. in the case of the $\gamma$-process where the models are typically not able to reproduce the observed p-isotope abundances. 
The problems of $\gamma$-process models triggered a huge experimental effort in the last decade aiming at the measurement of charged particle induced cross sections for testing the statistical model predictions. Although the experimental database is still somewhat limited and confined to the region of stable isotopes, the general observation is that statistical models strongly overestimate the experimental ($\alpha,\gamma$) cross sections of heavy isotopes. Deviations of up to an order of magnitude are found \cite{rau13}. 

Owing to the steeply falling cross section towards low energies, the cross sections are unfortunately not measured in the astrophysically relevant energy range, but above, where cross sections typically reach at least the $\mu$barn range. No direct information can thus be obtained from the measurements for the astrophysical processes, extrapolations are inevitable which involve serious difficulties.

The cross sections from statistical models are sensitive to various nuclear physics input parameters, like optical potentials, the $\gamma$-ray strength function, level densities, etc., which enter into the different reaction channel widths. Detailed studies show that the cross sections are not equally sensitive to the different widths and the sensitivities vary strongly with energy \cite{rau11}. In the case of $\alpha$-induced reactions at low, astrophysical energies the cross sections are only sensitive to the $\alpha$-width as this width is by far the smallest owing to the Coulomb barrier penetration. At higher energies, where $\gamma$-process related experimental $\alpha$-capture cross sections are available, however, the calculations are typically also sensitive to other widths. The simple comparison of the experimental results with model calculations therefore cannot reveal alone the incorrect nuclear input parameter. The study of ($\alpha$,n) reactions may help as the cross section of these reactions are usually sensitive only to the $\alpha$-width \cite[e.g.]{net13,kis14}. The probed energy range above the neutron threshold, however, is typically much higher than the astrophysically relevant one.

In spite of the fact that not the right energy range is probed, modifications of the $\alpha$-width obtained by the modification of the $\alpha$-nucleus optical potential are used for correcting the discrepancies between the measured and calculated ($\alpha,\gamma$) cross sections. The optical potential is considered to be the most uncertain, and therefore the key quantity in $\gamma$-process network calculations. Several different global $\alpha$-nucleus optical potential parameterizations are available and these potentials are continuously improved based on new experimental data. In spite of these efforts, however, there is still no global $\alpha$-nucleus optical potential which could describe the available experimental data of $\gamma$-process relevance. The study of the optical potential directly at astrophysical energies would therefore be highly needed but was not possible so far using the conventional method of studying $\alpha$-induced reactions.

Besides the optical potentials, an alternative solution for the discrepancies of the measured and calculated ($\alpha,\gamma$) cross sections has been suggested recently \cite{rau13b}. This approach considers direct reactions channels (like Coulomb excitation) which are not accounted for correctly in statistical model calculations. The argumentation suggests that the optical potential is actually correct, but part of the incoming $\alpha$-flux is removed by direct reaction channels and therefore the final cross section becomes lower than without the inclusion of this channel. One way of examining this possibility is to study the $\alpha$-nucleus optical potential in a reaction where the $\alpha$-particle is not in the entrance channel and hence Coulomb excitation (e.g.) cannot play a role.


\begin{figure}
\includegraphics[angle=270,width=0.8\columnwidth]{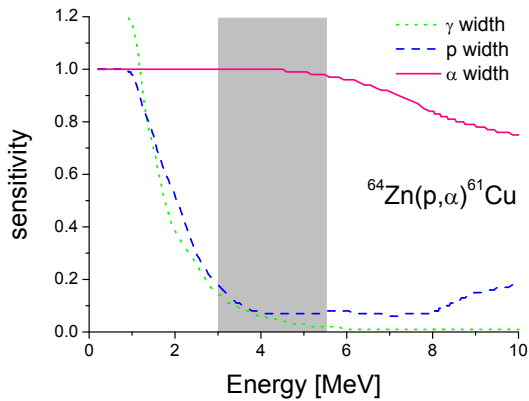} \\
\caption{\label{fig:sensitivity} (Color online) Sensitivity of the calculated $^{64}$Zn(p,$\alpha$)$^{61}$Cu cross section to the variation of the various partial widths. The shaded area shows the astrophysically relevant energy window for the inverse $^{61}$Cu\,+\,$\alpha$ system. See text for details.}
\end{figure}

Here we present the measurement of a (p,$\alpha$) reaction cross section for the first time in relation of heavy element nucleosynthesis. The $^{64}$Zn(p,$\alpha$)$^{61}$Cu reaction has been chosen for this purpose which has various advantages. First, this reaction is ideal for studying unambiguously the low energy $\alpha$-nucleus optical potential. Fig.\,\ref{fig:sensitivity} shows the sensitivities of the calculated $^{64}$Zn(p,$\alpha$)$^{61}$Cu cross sections to various partial widths. For the precise definition of the sensitivity see Ref.\,\cite{rau12}. Shortly, the sensitivity measures the change of the resulting cross section when a given width is changed. Zero sensitivity means the cross section does not change at all if a width is modified by a factor of two, while a sensitivity of one means that the cross section changes by the same factor as the width (full sensitivity). 

As one can see in the figure, the $^{64}$Zn(p,$\alpha$)$^{61}$Cu cross section is solely sensitive to the $\alpha$-width in the 3-8\,MeV energy range and here it shows a full sensitivity. Measuring the $^{64}$Zn(p,$\alpha$)$^{61}$Cu cross section in this energy range provides therefore direct information about the $\alpha$-width and thus for the $\alpha$-nucleus optical potential without any complication caused by Coulomb excitation.

Moreover, the information can be obtained directly at energies of astrophysical relevance. The astrophysically relevant energy range (Gamow window) for the inverse $^{61}$Cu\,+\,$\alpha$ system is between 3.8 and 6.5 MeV for a temperature of 3.5\,GK \cite{rau10} relevant for the $\gamma$-process in the lower mass range \cite{rau13}. Taking into account the $^{64}$Zn(p,$\alpha$)$^{61}$Cu reaction Q value of 844\,keV, this energy window translates into an energy range of about 3.0 -- 5.7 MeV for the $^{64}$Zn\,+\,p process studied in the present work. This energy range is shown in Fig.\,\ref{fig:sensitivity} as the gray shaded area. 

Consequently, by measuring the $^{64}$Zn(p,$\alpha$)$^{61}$Cu cross section and comparing the result with the predictions of statistical models, information can be obtained unambiguously for the $\alpha$-nucleus optical potential directly at astrophysical energies. Moreover, this is the first time when the optical potential is studied in the case of an unstable nucleus in relation to heavy element nucleosynthesis. No experimental data is available for this reaction at all in the literature and thus the aim of the present work was to measure this cross section in the energy range where the cross section is only sensitive to the $\alpha$-width as described above.

The second advantage of the $^{64}$Zn(p,$\alpha$)$^{61}$Cu reaction is that its reaction product is radioactive and therefore the well established activation method can be used for the cross section determination. The cross section was hence measured in the proton energy range between 3.5 and 8 MeV using the activation method. In this energy range the only other open reaction channel is the radiative capture (the neutron threshod is at 8.1\,MeV). Since this $^{64}$Zn(p,$\gamma$)$^{65}$Ga reaction also leads 
to a radioactive isotope, its cross section can also be determined with activation. Since in this energy range no experimental data exist for $^{64}$Zn(p,$\gamma$)$^{65}$Ga reaction either, as a side result of the present work this cross section was also measured. 

\begin{table}
\caption{\label{tab:decay} Decay parameters of the reaction products. Only the strongest gamma transitions used for the analysis are listed. Data are taken from \cite{NDS61} and \cite{NDS65}.
}
\begin{footnotesize}
\begin{ruledtabular}
\begin{tabular}{llll}
\hline
Reaction  & half-life & E$_\gamma$ & relative \\
          & 					 &	[keV]			&	intensity [\%] \\
\hline
$^{64}$Zn(p,$\alpha$)$^{61}$Cu & 3.33\,h &	283	&	12.2	$\pm$	0.3	$\pm$	2.2	\\
&	&	373	&	2.15	$\pm$	0.05	$\pm$	0.39	\\
&	&	589	&	1.17	$\pm$	0.02	$\pm$	0.21	\\
&	&	656	&	10.8	$\pm$	0.2	$\pm$	1.9	\\
&	&	909	&	1.10	$\pm$	0.02	$\pm$	0.20	\\
&	&	1185	&	3.75	$\pm$	0.07	$\pm$	0.68	\\
$^{64}$Zn(p,$\gamma$)$^{65}$Ga & 15.2\,min &	115	&	54.0	$\pm$	8.1	$\pm$	10.0	\\
&	&	153	&	8.9	$\pm$	0.9	$\pm$	1.6	\\
&	&	752	&	8.1	$\pm$	0.5	$\pm$	1.5	\\
\end{tabular}
\end{ruledtabular}
\end{footnotesize}
\end{table}

Table\,\ref{tab:decay} shows the decay parameters of the two reaction products. Only those $\gamma$-transitions are listed which were used for the analysis. It should be noted that in the case of both $^{61}$Cu  and $^{65}$Ga produced isotopes the normalization values of the relative $\gamma$-intensities have unusually high uncertainties of 18.0\,\% \cite{NDS61} and 18.5\,\% \cite{NDS65}, respectively. This is shown as the second uncertainty in the last column in table\,\ref{tab:decay}. These uncertainties represent by far the dominant error in the cross sections determined in the present work. 

The measurements were carried out using the experimental techniques described elsewhere \cite{gyu12}. Shortly, the cyclotron accelerator of Atomki provided proton beams in the energy range between 3.5 and 8\,MeV with typical beam intensities of about 1\,$\mu$A. The proton beam bombarded thin Zn targets enriched to 99.71\,\% in $^{64}$Zn. The targets were prepared by vacuum evaporation onto 2\,$\mu$m thick Al foils and their thicknesses were determined by weighing and Rutherford Backscattering Spectrometry. The lengths of the irradiations varied between 0.5 and 12 hours. The number of projectiles were determined by charge integration using multichannel scaling with one minute dwell time in order to take into account the variation of the beam intensity during the activations.

\begin{figure}
\includegraphics[angle=270,width=0.8\columnwidth]{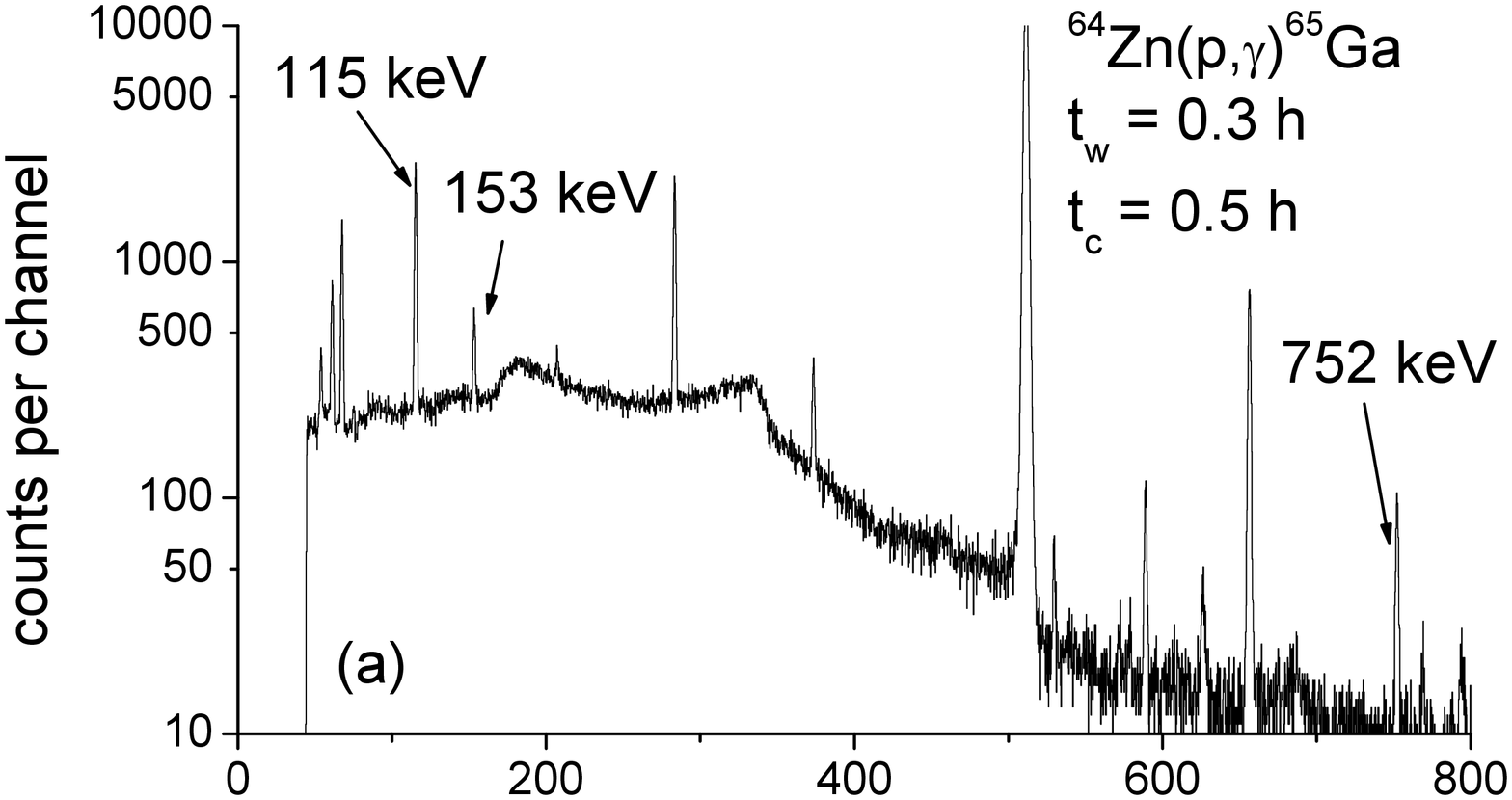} \\
\includegraphics[angle=270,width=0.8\columnwidth]{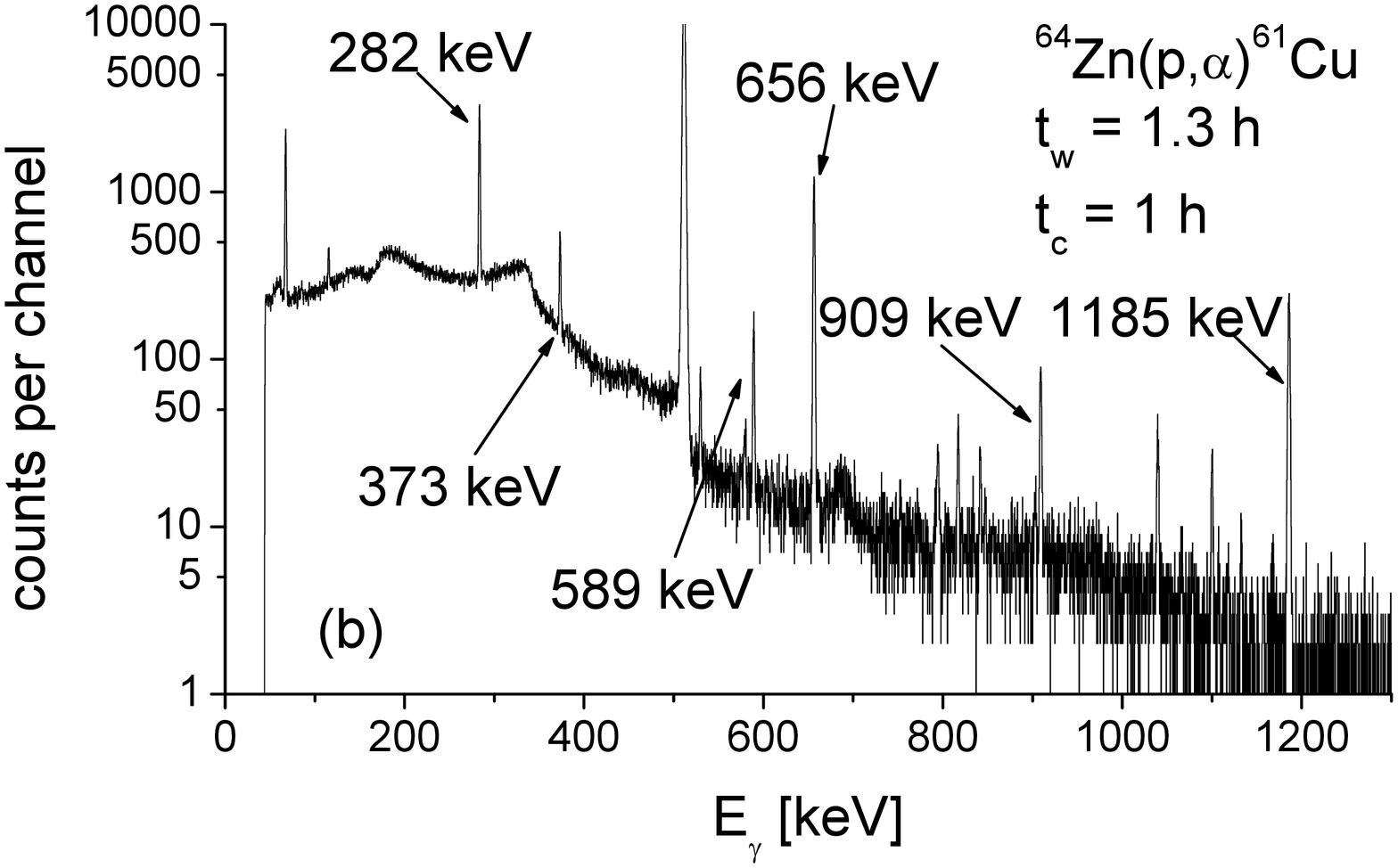}%
\caption{\label{fig:spectra} Activation $\gamma$-spectra measured on a $^{64}$Zn target irradiated with a proton beam of 7\,MeV. The upper panel (a) shows the first part of the counting where the spectrum is dominated by the decay of the short-lived $^{64}$Zn(p,$\gamma$)$^{65}$Ga reaction product, while the lower panel (b) shows a spectrum with the decay lines of the longer-lived $^{64}$Zn(p,$\alpha$)$^{61}$Cu reaction product (see the indicated waiting and counting times, t$\rm _w$ and t$\rm _c$, respectively). The peaks of the $\gamma$-transitions used for the analysis are marked.}
\end{figure}

The induced $\gamma$-activity was measured with a calibrated 100\,\% relative intensity HPGe detector equipped with complete 4$\pi$ low background shielding. Owing to the different half-lives of the two reaction products the $\gamma$-spectra measured in the first hour was used for the $^{64}$Zn(p,$\gamma$)$^{65}$Ga cross section determination, while the cross section of $^{64}$Zn(p,$\alpha$)$^{61}$Cu was obtained from the spectra taken afterwards. Typical $\gamma$-spectra after an irradiation at E$_p$\,=\,7\,MeV are shown in Fig.\,\ref{fig:spectra} separately for the two counting intervals.

\begin{figure}
\includegraphics[angle=270,width=0.8\columnwidth]{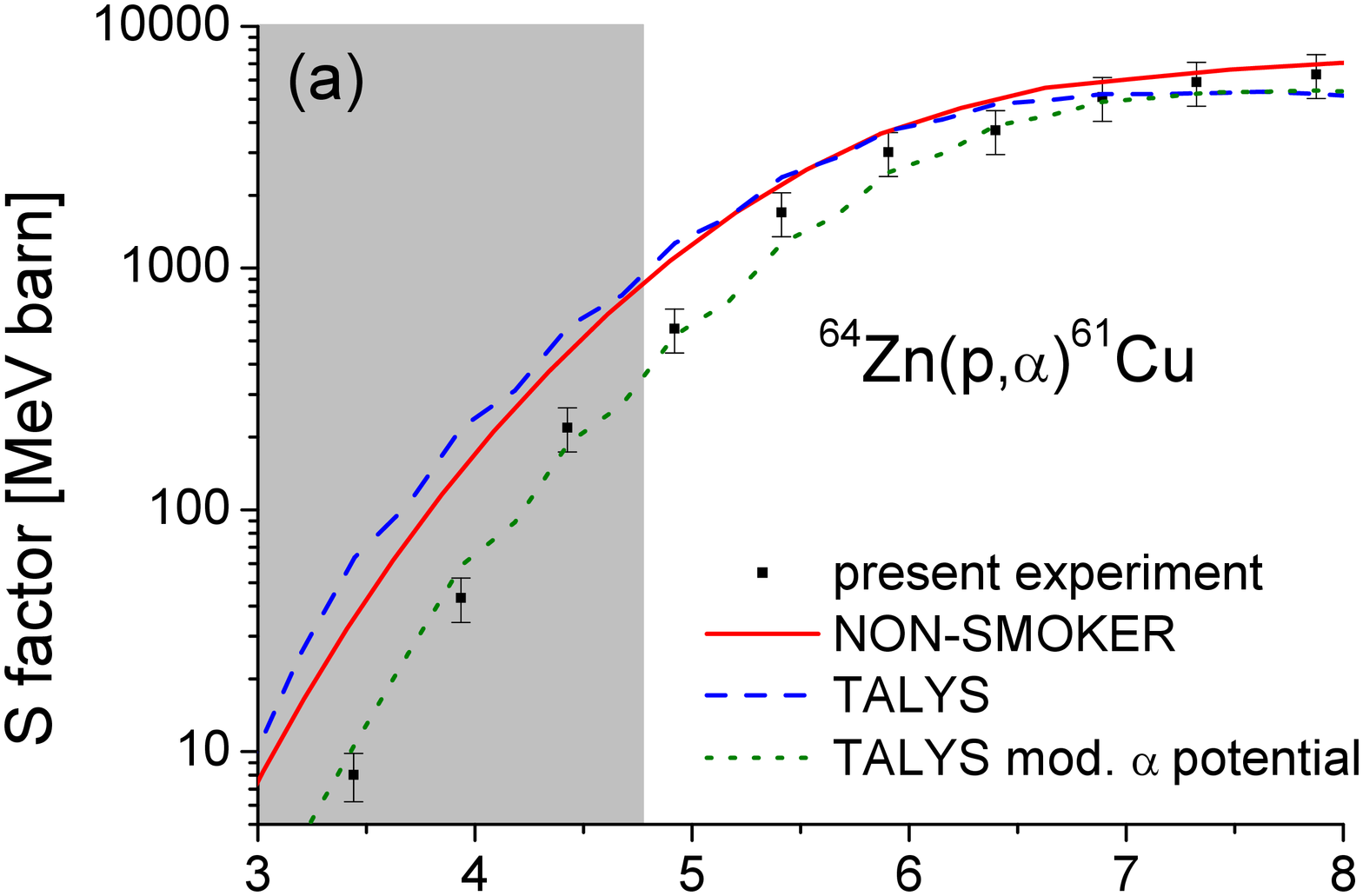} \\
\includegraphics[angle=270,width=0.8\columnwidth]{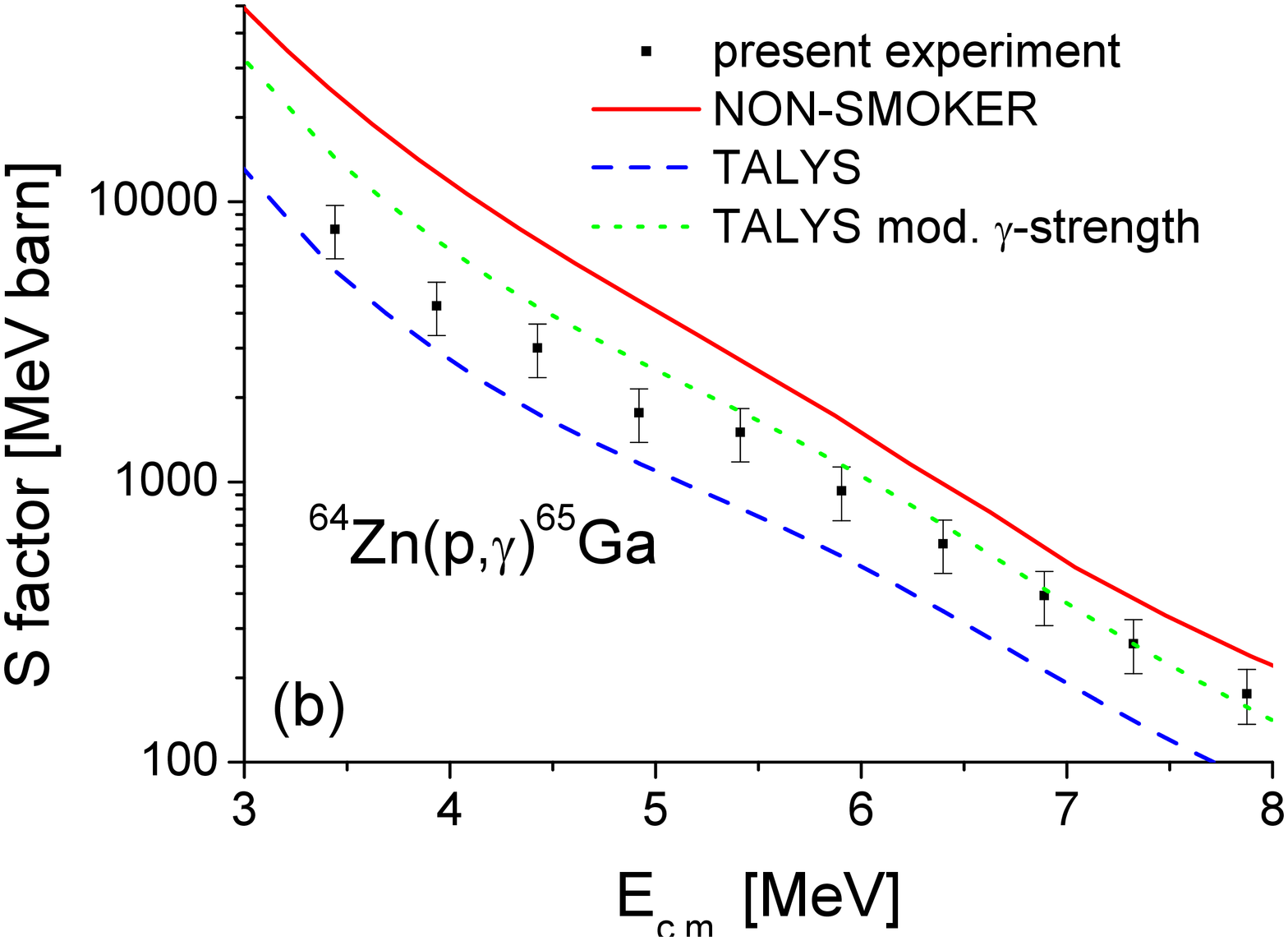}%
\caption{\label{fig:results} (Color online) Experimental S-factor of the $^{64}$Zn(p,$\alpha$)$^{61}$Cu (upper panel) and $^{64}$Zn(p,$\gamma$)$^{65}$Ga (lower panel) reactions and the predictions of statistical model calculations using the standard settings of the codes and a modified $\alpha$-nucleus optical potential and $\gamma$-ray strength function, respectively. The Gamow window for the inverse $^{61}$Cu\,+\,$\alpha$ system is shown again as a shaded area.}
\end{figure}

The measured cross sections are listed in table\,\ref{tab:results} and shown in Fig.\,\ref{fig:results} in the form of astrophysical S-factor. The uncertainty of the c.m. energies comes mainly from the beam energy calibration of the cyclotron. The most important sources of cross section uncertainty are the above mentioned normalization uncertainty of the relative $\gamma$-intensities (18.0\,\% and 18.5\,\%), target thickness determination (8\,\%), detection efficiency (5\,\%), charge collection (3\,\%), decay parameters ($<$5\,\%) and counting statistics ($<$10\,\%). 

The figures also show the results of the statistical model calculations carried out with the TALYS \cite{TALYS} and NON-SMOKER \cite{NON-SMOKER} codes. The predictions of the latter code is extensively used in astrophysical network calculations and therefore its comparison with experiments has important astrophysical consequences. In the case of the $^{64}$Zn(p,$\gamma$)$^{65}$Ga capture reaction one of the codes overestimate while the other underestimate the measured cross sections in the whole energy range. In this energy range the radiative capture cross section is mainly sensitive to the $\gamma$-width and therefore the deviation points to e.g. a deficiency in the $\gamma$-ray strength function.  Here the result of the TALYS calculations are also shown using a different $\gamma$-ray strength function, that of S. Goriely \cite{gor98}, which gives a somewhat better description of the present data then the standard strength of J. Kopecky and M. Uhl \cite{kop90}. The detailed discussion of the $^{64}$Zn(p,$\gamma$)$^{65}$Ga channel will be the subject of a forthcoming publication.

In the focus of the present paper is the study of the $\alpha$-nucleus optical potential through the $^{64}$Zn(p,$\alpha$)$^{61}$Cu reaction. The statistical models give a good reproduction of the measured data at the highest energies while start to deviate strongly towards lower, astrophysical energies (shaded area). At the lowest points the deviation reaches a value of a factor of about five to ten. This result provides the first direct evidence that at astrophysically relevant energies the statistical models (like NON-SMOKER using the standard optical potential of McFadden and Satchler \cite{McF66}) do not yield correct cross sections. 

The statistical model calculations were also carried out using different global $\alpha$-nucleus optical potentials. For this purpose the built-in potential of TALYS were used. The only good description of the experimental data was obtained with the potential of Demetriou \textit{et al.} using their dispersive model \cite{dem02}. This calculation is also shown in Fig.\,\ref{fig:results}. (It it worth noting that the modification of the $\gamma$-ray strength function as described above leads to the same cross section, within about 2\,\%. This supports the fact that the (p,$\alpha$) channel is only sensitive to the $\alpha$-width.)

\begin{table}
\caption{\label{tab:results} Measured cross sections of the two studied reactions.}
\begin{tabular}{r@{\hspace{2mm}}cl@{\hspace{8mm}}rcl@{\hspace{8mm}}rcl}
\hline
\multicolumn{3}{c}{E$^{\rm eff}_{\rm c.m.}$}& \multicolumn{3}{c}{$^{64}$Zn(p,$\alpha$)$^{61}$Cu}& \multicolumn{3}{c}{$^{64}$Zn(p,$\gamma$)$^{65}$Ga}\\
& & & \multicolumn{3}{c}{cross section}& \multicolumn{3}{c}{cross section}\\
\multicolumn{3}{c}{[MeV]}& \multicolumn{3}{c}{[$\mu$barn]}& \multicolumn{3}{c}{[$\mu$barn]}\\
\hline
3.44	&	$\pm$	&	0.03	&	0.333	&	$\pm$	&	0.075	&	331	&	$\pm$	&	71	\\
3.94	&	$\pm$	&	0.04	&	4.34	&	$\pm$	&	0.90	&	427	&	$\pm$	&	92	\\
4.43	&	$\pm$	&	0.04	&	45.4	&	$\pm$	&	9.4	&	624	&	$\pm$	&	135	\\
4.92	&	$\pm$	&	0.05	&	213	&	$\pm$	&	44	&	672	&	$\pm$	&	145	\\
5.41	&	$\pm$	&	0.05	&	1085	&	$\pm$	&	223	&	960	&	$\pm$	&	207	\\
5.90	&	$\pm$	&	0.06	&	3013	&	$\pm$	&	621	&	928	&	$\pm$	&	201	\\
6.40	&	$\pm$	&	0.06	&	5493	&	$\pm$	&	1131	&	889	&	$\pm$	&	192	\\
6.89	&	$\pm$	&	0.07	&	10651	&	$\pm$	&	2197	&	822	&	$\pm$	&	180	\\
7.32	&	$\pm$	&	0.08	&	16163	&	$\pm$	&	3328	&	727	&	$\pm$	&	159	\\
7.88	&	$\pm$	&	0.08	&	23746	&	$\pm$	&	4894	&	658	&	$\pm$	&	146	\\
\hline
\end{tabular}
\end{table}

The largely different cross sections predicted with various optical potentials may have strong astrophysical consequences. The astrophysical reaction rate of the $^{61}$Cu($\alpha,\gamma$)$^{65}$Ga reaction has been calculated with TALYS using the McFadden and Satchler \cite{McF66} and Demetriou \textit{et al.} potentials \cite{dem02}. The two rates differ by a factor of five at 3.5\,GK $\gamma$-process temperature, while the deviation goes up to one order of magnitude at 2\,GK. Since the first potential is used in many astrophysical network calculations, while the second one gives a good description of the present experimental data, direct experimental evidence is provided for the strongly overestimated reaction rate of the $\gamma$-process network calculations in the case of $^{61}$Cu($\alpha,\gamma$)$^{65}$Ga.

The result of the present work provides a direct evidence of an incorrect optical potential only in the case of the $^{61}$Cu\,+\,$\alpha$ system at astrophysical energies. If, on the other hand, one takes into account the general observation, that at higher energies the standard global optical potentials lead to too high cross sections, one can conclude that $\gamma$-process models in general use strongly overestimated rates for reactions involving $\alpha$-particles. This can have strong consequences for the prediction of p-isotope abundances. In order to put the conclusion of the present work on a more solid ground, further direct experimental study of the $\alpha$-nucleus optical potential at astrophysical energies is required. The further application of (p,$\alpha$) reactions introduced in this work may contribute to this aim.

\begin{acknowledgments}
This work was supported by OTKA (K101328, PD104664, K108459). G.G. Kiss acknowledges support from the J\'anos Bolyai Research Scholarship of the Hungarian Academy of Sciences.
\end{acknowledgments}

\end{document}